\documentclass[aps,twocolumn]{revtex4}

\newcommand{\ww}{\mbox{\tiny $\wedge$}}

\begin{document}

\title{The linear spectrum of $OSp(32|1)$ Chern-Simons supergravity in eleven
dimensions}

\author{M\'aximo Ba\~nados}

\affiliation{Departamento de F\'{\i}sica, P. Universidad Cat\'olica de Chile, Casilla 306, Santiago 22,
Chile.
\\ {\tt mbanados@fis.puc.cl}}

\begin{abstract}

We study linearized perturbations of eleven-dimensional $OSp(32|1)$ 
Chern-Simons supergravity.  The action contains a term that changes the value of the cosmological constant, as considered by Horava. It is shown that the spectrum contains a 3-form and a 6-form whose field strengths are dual to each other, thus providing a link with the eleven-dimensional supergravity of Cremmer, Julia and Scherk. The linearized equations for the graviton and Rarita-Schwinger field are shown to be the standard ones as well. 


\end{abstract}

\maketitle

Eleven-dimensional supergravity \cite{Cremmer-JS} has become widely accepted as one of the key stones towards the Theory of Everthing. 
This theory represents the low energy regime of M-theory which contains, in varius limits, all string theories. Moreover, after compactification, string theory is expected to contain a phenomenologically correct model of our Universe. (See \cite{Duff} for a complete list of references and reprinted papers on this subject.)

It has been known for a long time that the underlying symmetry group of
eleven-dimensional supergravity should be the ortosymplectic group $OSp(32|1)$. This was already conjectured in the original paper by Cremer,Julia and Scherk \cite{Cremmer-JS}, and it has recently become important in the context of the M-
theory supersymmetry algebra
\cite{Townsend,Dijkgraaf-VV},
\begin{equation}
\{Q,\bar Q \} = \gamma^a P_a + \gamma^{ab}
Z_{ab} + \gamma^{a_1...a_5} Z_{a_1...a_5}.
\label{M}
\end{equation}
$Z_{ab}$ and $Z_{a_1...a_5}$ are central charges corresponding to the
two types of extended objects, M2 and M5, that couple to the three form.  However,
the action constructed in \cite{Cremmer-JS,Nahm} is not invariant under $OSp(32|1)$, at
least not in any obvious way.  See
\cite{Castellani-FGPvN,Nicolai-TvN,Auria-F} for
discussions on eleven dimensional supergravity and $OSp(32|1)$, and
\cite{Horava,Bandos-BS,Nishino99,Ling-S} for other related aspects.

There exists another supergravity theory in eleven dimensions constructed as a
Chern-Simons density for $OSp(32|1)$ \cite{TZ,BTrZ}. This action, which
is a direct generalization of the well-known three-dimensional case
\cite{Achucarro-TW}, is explicitly invariant under this supergroup. The equations
of motion are however more complicated and there are no known solutions other
than some simple examples with spherical symmetry
\cite{BTZ2}.

It is desirable to have a relation between these two theories specially given the key role played by standard supergravity in string and M-theory. Horava \cite{Horava} has conjectured  that this relation exists (see also \cite{TZ}). In particular, it is claimed in that reference that the three-form of Cremmer, Julia and Scherk should be contained in the $OSp(32|1)$ field.   

It is the goal of this paper to put this conjecture on a more firm ground. We shall prove that the equations considered by Horava do indeed contain a three-form satisfying the correct equations, at least at the linearized level. It will also follow from the analysis that the graviton and Rarita-Schwinger fields satisfy the correct linear equations.   

The main input in the Chern-Simons construction is the $OSp(32|1)$ gauge field
carrying one field for each generator in (\ref{M})
\begin{equation}
W =  { 1 \over 2l}\, e^a \gamma_a + {1 \over 4}\, w^{ab} \gamma_{ab} +
{1 \over 5! }\, A^{a_1...a_5}\gamma_{a_1...a_5} + \psi \bar Q
\label{W}
\end{equation}
with a natural identification of $e^a$ as the vielbein, and $w^{ab}$ as the spin
connection. The constant $l$ is the AdS scale. In what follows we set the fermions
equal to zero.

Comparing (\ref{M}) and (\ref{W}) we are led to identify the `electric' charge $Z_{ab}$ with the spin connection (more precisely the torsion $T^a=De^a$) and the `magnetic' charge $Z_{a_1...a_5}$ with  $A^{a_1...a_5}$. In fact, let $A_{[3]}$ and $A_{[6]}$ be the totally antisymmetric parts of $T^a$ and $A^{a_1...a_5}$,
\begin{eqnarray}
A_{[3]} &=& {1\over 3} T_a \ww e^a, \\ 
A_{[6]} &=& {1 \over 6!} \, A_{a_1...a_5}\ww e^{a_1}\ww ...\ww e^{a_5}. \label{A6}
\end{eqnarray}
We shall prove that the linearized equations following from a Chern-Simons theory, yield the expected duality relation,
\begin{eqnarray}
 dA_{[3]} &=& \alpha^* \! dA_{[6]} \label{dual}
\end{eqnarray}
where $\alpha = 1/50$.  See \cite{Bandos-BS,Nishino99} for some recent work on a `duality-symmetric' formulation of eleven dimensional supergravity. (An interacting supergravity theory containing only $A_{[6]}$ does not seem to exist \cite{Nicolai-TvN}.) 

Let $\hat R=dW + W  W$, the Chern-Simons Lagrangian ${\cal L}_{CS}$ is defined as
Tr$\hat R^6 = d {\cal L}_{CS}$. The explicit expression for ${\cal L}_{CS}$ can be
found in \cite{Zumino83} and the corresponding formula for $OSp(32|1)$ in
\cite{TZ}.
The Chern-Simons action in eleven dimensions is
\begin{equation}
I_{CS} = k\int_{M_{11}} {\cal L}_{CS}.
\label{ICS}
\end{equation}
A quantization condition for $k$ has been discussed in \cite{JZ}. The
equations of motion following from (\ref{ICS}) are 
\begin{equation}
\hat R^5=0.
\label{CSeq}
\end{equation}
The curvature $\hat R=dW + W\ww W$ expanded in the basis
$\{\gamma_a,\gamma_{ab},\gamma_{a_1...a_5}\}$ is
\begin{equation}
\hat R = {1 \over 2l} T^a \gamma_a + {1 \over 4} \bar R^{ab} \gamma_{ab}
+ {1 \over 5!} \bar F^{a_1...a_5} \gamma_{a_1...a_5}
\label{hatR}
\end{equation}
where  $T^a$ is the torsion, $\bar R^{ab} = R^{ab} +  e^a \ww e^b / l^2$ and
\begin{equation}
\bar F^{a_1...a_5} = DA^{a_1...a_5} + {1 \over 5!\, l}
\epsilon^{a_1...a_5}_{\ \ \ \ \ \ \  b_1...b_6}  e^{b_1}\ww A^{b_2...b_6}.
\nonumber
\end{equation}
Here we have kept only linear terms in $A^{a_1...a_5}$.  $R^{ab}$ and $D$ are the
Lorentz curvature and covariant derivative.

If the Lie algebra is taken as $SO(10,2) \subset Sp(32)$ ($A^{a_1...a_5}=0$), then (\ref{ICS}) represents a pure theory of gravity in eleven dimensions of the Lovelock type \cite{Lovelock}, with all coefficients fixed in a particular way \cite{Chamseddine,BTZ2}.

Despite the formal similarities of this construction with the three-dimensional case \cite{Achucarro-TW}, there are important differences. First, the theory following from (\ref{ICS}) has non-trivial dynamics with a non-zero number of degrees of freedom \cite{BGH}. In fact, contrary to the three-dimensional case, propagating solutions are expected to exist although, to our knowledge, have not been found so far (the analysis of \cite{BGH} is restricted to the counting of independent initial conditions per spacetime point; no explicit solutions are displayed). Second, the action (\ref{ICS}) is {\it not} equivalent to the Einstein-Hilbert action in eleven dimensions. As mentioned before,  writting (\ref{ICS}) as a function of the spacetime curvature one obtains a Lovelock-type \cite{Lovelock} action containing higher powers of the curvature tensor.       

Equations (\ref{CSeq}) have a large group of symmetries $Sp(32)\times$diffs.  The maximally symmetric background $\hat R=0$ (AdS space) is however degenerate and no linear theory exists around it.	 Perturbations with spherical symmetry, for example, do not decay as $\delta g_{00} \sim 2M/r^8$ \cite{BTZ2}. This problem can be corrected if one breaks part of the gauge symmetry.  We shall do this by adding to the action a cosmological term which preserves Lorentz and diffeomorphism invariance. This simple modification produces asymptotically Schwarzschild spacetimes \cite{Boulware-D,Wheeler,5d}, and we show in this paper that an interesting linear theory exists around the new AdS background. Interestingly, a similar modification was considered in \cite{Horava} in an attempt to relate Chern-Simons and standard supergravities. Further implications of this term will be discussed in the conclusions.

We consider the deformed Chern-Simons action
\begin{equation}
I_\tau= I_{CS} - {k\, \tau^5 \over l^{11}} \int \mbox{Tr} \ e^{a_1}\ww \cdots \ww
e^{a_{11}}
\gamma_{a_1...a_{11}}
\label{tau}
\end{equation}
where $\tau$ is a dimensionless number parametrizing the deformation.  The new piece
is of course equivalent to a cosmological term. The equations of motion
following from $I_\tau$ are
\begin{equation}
\hat R^5 - {\tau^5 \over l^{10}} e^{a_1} \ww \cdots \ww e^{a_{10}}
\gamma_{a_1...a_{10}} =0.
\label{eq2}
\end{equation}

The new background is then
\begin{equation}
\bar R^{ab} = {4\tau \over l^2} e^a\ww e^b, \ \ \ A^{a_1...a_5}=T^{a}=0
\label{newbg}
\end{equation}
and represents a space of constant curvature with a modified cosmological
constant $$ \Lambda = {-1 + 4\tau  \over l^2}.$$

In view of the theorem proved in \cite{Bautier-DHS}, we expect to make contact with standard supergravity only in the limit $\Lambda=0$. In fact, we shall see that, for any value of $\tau$, the scale $l$ induces a mass term $m=1/l$ on the 3-form $A_{[3]}$ (see below). For this reason, in what follows we shall consider the $l\rightarrow \infty$ limit.

The linearized perturbations on $e^a$ and $\psi$ around the new background
(\ref{newbg}) have no surprises, so we only quote the result. Let $T^a=
A^{a_1...a_5}=0$ and we perturb the vielbein as $e^a + \delta e^a$ where $e^a$ is the
solution of (\ref{newbg}). Eq. (\ref{eq2}) becomes (in the large $l$ limit)
\begin{equation}
\epsilon_{a_1... a_{11}} \delta R^{a_1 a_2}\ww e^{a_3} \ww \cdots \ww e^{a_{10}}
=0
\label{grav}
\end{equation}
which is the linearized Einstein equation in eleven dimensions.
On the other hand, the linearized Chern-Simons fermion equation is
\cite{TZ}
\begin{equation}
\bar R^{a_1 a_2} \ww \cdots \ww \bar R^{a_7 a_8}  \gamma_{a_1...a_8} D \psi =
0.
\label{fermi}
\end{equation}
On the original background $\bar R^{ab}=0$ this equation is trivial. On the new
background (\ref{newbg}) we find the standard Rarita-Schwinger
equation $\tau^4 e^{a_1}\ww \cdots \ww e^{a_8} \gamma_{a_1...a_8} D\psi =0$ in
eleven dimensions.  (In the finite $l$ case (\ref{grav}) has a cosmological
term, and (\ref{fermi}) a mass term.)

The linearized equations for $A^{a_1...a_5}$ and $T^a$ can be computed as follows.
The curvature $\hat R$ expanded to first order in $A^{a_1...a_5}$ and $T^a$ is 
$\hat R = (\tau / l^2) e^a\ww e^b \gamma_{ab} + H$ where 
\begin{equation}
H = {1 \over 2l} T^a \gamma_a+{1 \over  4} D \kappa^{ab} \gamma_{ab} + {1 \over 5!}  \bar F^{a_1...a_5} \gamma_{a_1...a_5}.
\end{equation}
Here we have split the spin connection as $w^{ab} = w^{ab}(e) + \kappa^{ab}$ where $w^{ab}(e)$ is the torsionless part. $\kappa^{ab}$ is related to the torsion as $T^a = \kappa^a _{\ b} \ww e^b$.  The equations of motion (\ref{eq2}), keeping only linear terms
in $H$, can be written in the convenient form 
\footnote{$H$ and $e^a\ww e^b \gamma_{ab}$ are both 2-forms with values on the 32$\times$32 matrix space. Let $\gamma_n = e^{a_1} \ww \cdots \ww
e^{a_n} \gamma_{a_1...a_n}$ with $\gamma_n \ww \gamma_m =\gamma_{n+m}$.
Expanding $(\gamma_2+ H) ^5-\gamma_{10}=0$ to first order in $H$   yields
$ \gamma_8 \ww H + \gamma_6 \ww H \ww \gamma_2 + \gamma_4\ww H\ww
\gamma_4 + \gamma_2 \ww H \ww \gamma_6 + H \ww \gamma_8 = (1/2)\{\{\{H,\gamma_2\},\gamma_4\},\gamma_2\} +(1/2)\{H,\gamma_8\}= 0$.}
\begin{eqnarray}
 e^{a_1} \ww \cdots \ww e^{a_8} \ww \{\{\{H,\gamma_{a_1 a_2}\},\gamma_{a_3 a_4
a_5 a_6}\},\gamma_{a_7 a_8}\}  + & \nonumber \\   e^{a_1} \ww \cdots \ww e^{a_8}
\ww \{H,\gamma_{a_1...a_8}\} =& 0. \nonumber
\end{eqnarray}
Inserting $H$ and computing the anti-commutators \cite{Gamma} one finds an equation of the
form $\varepsilon^a \gamma_a+ \varepsilon^{ab} \gamma_{ab} + \varepsilon^{a_1...a_5}
\gamma_{a_1...a_5}=0$ where $\varepsilon^a=0$, $\varepsilon^{ab}=0$ and $\varepsilon^{a_1...a_5}=0$ are, respectively,
\begin{eqnarray}
\epsilon_{a_1...a_{11}} e^{a_1}\ww \cdots \ww e^{a_8}\ww D \kappa^{a_9 a_{10}} &= 0 \nonumber\\
\epsilon_{a_1...a_{11}} e^{a_1}\ww \cdots \ww e^{a_6} \ww (e^{a_7}\ww e^{a_8} \ww T^{a_9} -2l \bar F^{a_7a_8 a_9})  &= 0  \nonumber \\
e^{[a_1} \ww \cdots \ww e^{a_4} \ww \bar F^{a_5]} + {50 \over 7} \epsilon^{a_1...a_{11}} e_{a_6} \cdots \ww e_{a_{11}} dA_{[3]}  &= 0 \nonumber 
\end{eqnarray}
Here we have defined $\bar F^a= \bar F^a_{\ a_1...a_4} \ww e^{a_1} \cdots \ww e^{a_4}$ and $\bar F^{abc}= \bar F^{abc}_{\ \ \ a_1a_2} \ww e^{a_1} \ww e^{a_2}$

Eliminating the tangent indices, the above equations can be written in a more useful form   
\begin{eqnarray}
d^* \! A_{[3]} &=& 0 \label{Lorentz}\\
{7 \over 3l} \,  A_{[3]} &=&~ ^*\! dA_{[7]} \label{w2}   \\   
dA_{[6]} + {1 \over l}  A_{[7]} &=& 50\, ^* \!dA_{[3]}   \label{dual1}
\end{eqnarray}
with \footnote{Actually, (\ref{dual1}) is a projection of $\varepsilon^{a_1...a_5}=0$, namely $\varepsilon^{a_1...a_5}\ww e_{a_5}=0$. The other components involve the mixed tensor $S$ and the trace $C$, which is the dual of $A_{[7]}$, appearing in the irreducible decomposition   $A^{\mu_1...\mu_5 \ \mu} = A_{[6]}^{\ \mu_1...\mu_5  \mu} + C^{[\mu_1...\mu_4} \eta^{\mu_5] \mu} + S^{\mu_1...\mu_5 \ \mu}$.   
 The tensor $S$ is traceless and satisfies a Jacobi identity $S^{[\mu_1...\mu_5 \ \mu]}=0$.  Mixed tensors of this type have appeared in dual formulations of General Relativity in \cite{Curtright,Urrutia,Hull}. We shall consider these equations in detail elsewhere. The important point is that they do not impose further restrictions on $A_{[3]}$ and $A_{[6]}$.} 
\begin{equation}
A_{[7]} = {1 \over 6!\, 5!} \, \epsilon_{a_1...a_{11}} A^{a_1...a_5} \ww
e^{a_6}\ww \cdots \ww e^{a_{11}}
\end{equation}

Eq. (\ref{Lorentz}) is a Lorentz gauge condition for $A_{[3]}$. The field $A_{[7]}$ can be solved from (\ref{dual1}) and replacing in (\ref{w2}) we find that $A_{[3]}$ has a mass proportional to $1/l$, 
\begin{equation}
^* \! d ^*\! dA_{[3]} = {7 \over 150 \, l^2} A_{[3]}.
\end{equation}
In the limit  $l\rightarrow \infty$ this mass term disappears, we recover the usual dynamics for the three-form and (\ref{dual1}) becomes the duality relation (\ref{dual}).   

The appearance of a Lorentz gauge condition was somehow expected. The group of symmetries of the action, {\it Lorentz $\times$ diff.}, does not act on $A_{[3]}=(1/3) T_a\ww e^a$ as a gauge transformation. Hence, the description is not gauge invariant \footnote{It is interesting to note, however, that the truncated 3-form $A'_{[3]} =-(1/3) w_{ab}\ww e^a \ww e^b$, which is motivated by (\ref{M}) and (\ref{W}),  does transform as a gauge field under the Lorentz group.}.      

$OSp(32|1)$ Chern-Simons theory is then a topological (metric-independent)
and gauge invariant field theory whose linear spectrum (perturbations around a deformed maximally symmetric background) contains linearized eleven-dimensional supergravity. Since eleven-dimensional supergravity is itself the low energy regime of M-theory, it is tempting to conjecture (see also \cite{TZ,Horava}) a direct relation between Chern-Simons supergravity and M-theory. To make this relation precise the inclusion of interactions (higher order terms) is however necessary. We expect to come back to this problem elsewhere. 
 
It is perhaps not too surprising that in order to find a well-defined and
interesting linear theory we had to break the Chern-Simons symmetry. In field theory
often the interesting physics arises in broken phases for which some
fields take non-zero expectation values. It would be extremely interesting to find
an exact solution inducing the symmetry breaking term in a spontaneous way. Alternatively, one can understand this term via the coupling of Wilson lines, as in \cite{Horava}. 

It is also interesting to note that the symmetry breaking term only affects the
asymptotic behavior of the metric \cite{5d}.  Near the singularity, the physics is controlled
by the exact Chern-Simons theory which, in particular, leads to a milder
singularity \cite{BTZ2}.

The results presented here hold also in five dimensions. The actions of 5d supergravity 
\cite{Chamseddine-N,DAuria-MRF} and 5d Chern-Simons supergravity \cite{Chamseddine}
are not equivalent. It can be shown \cite{5d} that deforming the Chern-Simons action by the addition of a cosmological term, its linearized spectrum coincides with standard supergravity. 
One can also go to second order and see the Abelian Chern-Simons
interaction, as well as the back reaction from the gauge field. The equations of
motion of a spherically symmetric ansatz can be solved and the solution coincides,
asymptotically, with the usual five-dimensional Reissner-Nordstrom solution
\cite{5d}.

Finally, some open problems that we have left for future work are the expansion to 
second order (see \cite{5d} for the five-dimensional case), existence of other
backgrounds, like the M2/M5 solutions, and existence of dualities via dimensional
reduction \cite{Witten95,Berghshoeff-HO}. If these exist, then perhaps Chern-Simons
supergravity may become as relevant as standard supergravity. It would be
interesting to study the strong regime of this theory where the non-linear terms
become important (and the symmetry is restored) and see their effect on various
singularities.

~

The author would like to thank M. Henneaux for useful comments, and A. Gomberoff for
many useful and stimulating discussions. This work was partially supported by grant
\# 1000744
(FONDECYT,
Chile).


\begin{thebibliography}{10}


\bibitem{Cremmer-JS} E. Cremmer, B. Julia and J. Scherk
 Phys.Lett. {\bf B76}, 409 (1978)

\bibitem{Duff}
M.~J.~Duff,
{\it  Bristol, UK: IOP (1999) 513 p}.

 \bibitem{Townsend}
 P.~K.~Townsend,  hep-th/9507048; hep-th/9612121

\bibitem{Dijkgraaf-VV}
R.~Dijkgraaf, E.~Verlinde and H.~Verlinde,
Nucl.\ Phys.\ Proc.\ Suppl.\  {\bf 62}, 348 (1998)

\bibitem{Nahm}
W.~Nahm,  Nucl.\ Phys.\  {\bf B135}, 149 (1978).

\bibitem{Castellani-FGPvN}
L.~Castellani, P.~Fre, F.~Giani, K.~Pilch and P.~van Nieuwenhuizen,
Annals Phys.\  {\bf 146}, 35 (1983).

\bibitem{Nicolai-TvN}
H.~Nicolai, P.~K.~Townsend and P.~van Nieuwenhuizen,
Lett.\ Nuovo Cim.\  {\bf 30}, 315 (1981).

\bibitem{Auria-F}
R.~D'Auria and P.~Fre,
Nucl.\ Phys.\ B {\bf 201} (1982) 101
[Erratum-ibid.\ B {\bf 206} (1982) 496].

\bibitem{Horava}
P.~Horava,
Phys.\ Rev.\ {\bf D59}, 046004 (1999)
[hep-th/9712130].

\bibitem{Bandos-BS}
I.~Bandos, N.~Berkovits and D.~Sorokin,
Nucl.\ Phys.\ B {\bf 522}, 214 (1998)
[hep-th/9711055].

\bibitem{Nishino99}
H.~Nishino, Mod.\ Phys.\ Lett.\ A {\bf 14}, 977 (1999)
[hep-th/9802009].

\bibitem{Ling-S}
Y.~Ling and L.~Smolin,
Nucl.\ Phys.\ B {\bf 601}, 191 (2001)
[hep-th/0003285].

\bibitem{TZ} R. Troncoso and J. Zanelli,
Phys.Rev. {\bf D58}, R101703 (1998);
R.~Troncoso and J.~Zanelli, hep-th/9902003.

\bibitem{BTrZ} M. Ba\~nados, R. Troncoso and J.
        Zanelli,  Phys. Rev. {\bf D54}, 2605 (1996).

\bibitem{Achucarro-TW} A. Ach\'ucarro and
P.K. Townsend,  Phys. Lett. {\bf B180}, 89 (1986). E. Witten,  Nucl. Phys. {\bf B
311}, 4 (1988).

\bibitem{BTZ2} M. Ba\~nados, C. Teitelboim and J.Zanelli,
               Phys. Rev. {\bf D49}, 975 (1994).
               
\bibitem{Zumino83}
B.~Zumino, ``Chiral Anomalies And Differential Geometry", UCB-PTH-83/16
{\it Lectures given at Les Houches Summer School on Theoretical Physics, Les
Houches, France, Aug 8 - Sep 2, 1983}.
  
\bibitem{JZ}
J.~Zanelli, Phys.\ Rev.\ D {\bf 51}, 490 (1995)
[hep-th/9406202].
               
\bibitem{Lovelock} D. Lovelock,  J. Math. Phys. {\bf 12}, 498
(1971).
  
\bibitem{Chamseddine} A.H. Chamseddine,
Nucl. Phys. {\bf B346}, 213 (1990).           

\bibitem{BGH} M. Ba\~nados, L.J. Garay and M. Henneaux,
              Phys. Rev. {\bf D53}, R593 (1996);  Nucl. Phys. {\bf B476}, 611
              (1996).    
    
\bibitem{Boulware-D}
D.G.~Boulware and S.~Deser,
	       Phys.\ Rev.\ Lett.\ {\bf 55}, 2656 (1985).

\bibitem{Wheeler} J.T. Wheeler,  Nucl.Phys. {\bf B268}, 737
	          (1986); {\bf B273}, 732 (1986).
             
\bibitem{5d}  M. Ba\~nados, submitted to Physical Review D. See also
M.~Ba\~nados,  Nucl.\ Phys.\ Proc.\ Suppl.\  {\bf 88}, 17 (2000)
[hep-th/9911150].
 
\bibitem{Bautier-DHS}
K.~Bautier, S.~Deser, M.~Henneaux and D.~Seminara,
Phys.\ Lett.\ B {\bf 406}, 49 (1997)
[hep-th/9704131].

\bibitem{Gamma} See U. Gran, hep-th/0105086 for a useful Mathematica package to do
calculations with Dirac matrices.

\bibitem{Curtright}
T.~L.~Curtright and P.~G.~Freund,
Nucl.\ Phys.\ B {\bf 172}, 413 (1980).

\bibitem{Urrutia}
H.~Casini, R.~Montemayor and L.~F.~Urrutia,
Phys.\ Lett.\ B {\bf 507}, 336 (2001)
[hep-th/0102104].

\bibitem{Hull}
C.~M.~Hull,
hep-th/0107149.

\bibitem{DAuria-MRF}
R.~D'Auria, E.~Maina, T.~Regge and P.~Fre,
Annals Phys.\  {\bf 135} (1981) 237.

\bibitem{Chamseddine-N}
A.H.~Chamseddine and H.~Nicolai,
Phys.\ Lett.\ {\bf B96}, 89 (1980).

\bibitem{Witten95}
E.~Witten,
Nucl.\ Phys.\ B {\bf 443}, 85 (1995)
[hep-th/9503124].

\bibitem{Berghshoeff-HO}
E.~Bergshoeff, C.~Hull and T.~Ortin,
Nucl.\ Phys.\ B {\bf 451}, 547 (1995)
[hep-th/9504081].

\end{thebibliography}
\end{document}